\documentclass[conference]{IEEEtran}

\usepackage{cite}
\usepackage{amsmath,amssymb,amsfonts,braket}
\usepackage{algorithm}
\usepackage{tabularx}
\usepackage{booktabs}
\usepackage{comment}
\usepackage{float}
\usepackage{xspace}

\usepackage{quantikz}
\usepackage{tikz}
\usepackage{mathtools}
\usepackage{bm}
\usepackage{placeins}

\usepackage[top=0.75in, bottom=1.05in, left=0.63in, right=0.63in]{geometry}

\usepackage{xcolor}

\tikzset{
  node/.style = {draw, rounded corners, inner sep=6pt, font=\small},
  op/.style   = {font=\small\itshape, inner sep=0pt},
  arr/.style  = {-{Stealth}, line width=.6pt},
}

\DeclareUnicodeCharacter{2192}{\ifmmode\rightarrow\else{$\rightarrow$}\fi}
\DeclareUnicodeCharacter{2208}{\ifmmode\in\else{$\in$}\fi}
\DeclareUnicodeCharacter{00D7}{\ifmmode\times\else{$\times$}\fi}
\DeclareUnicodeCharacter{2218}{\ifmmode\circ\else{$\circ$}\fi}
\DeclareUnicodeCharacter{2265}{\ifmmode\geq\else{$\geq$}\fi}
\DeclareUnicodeCharacter{27E8}{\ifmmode\langle\else{$\langle$}\fi}
\DeclareUnicodeCharacter{2297}{\ifmmode\otimes\else{$\otimes$}\fi}
\DeclareUnicodeCharacter{27E9}{\ifmmode\rangle\else{$\rangle$}\fi}

\AtBeginDocument{%
  }

\begin{document}

\title{Quantum Machine Learning for Cybersecurity Applications: Simulation and Hardware Validation}

\author{\IEEEauthorblockN{Zirui Zhu}
\IEEEauthorblockA{\textit{Information Security Institute}\\ \textit{Johns Hopkins University} \\ Baltimore, USA\\
ziruizhu87@gmail.com}
\and
\IEEEauthorblockN{Zisheng Chen}
\IEEEauthorblockA{\textit{Information Security Institute}\\ \textit{Johns Hopkins University} \\ Baltimore, USA\\
zisheng.chen.2020@gmail.com}
\and
\IEEEauthorblockN{Xiangyang Li}
\IEEEauthorblockA{\textit{Information Security Institute}\\ \textit{Johns Hopkins University} \\ Baltimore, USA\\
xyli@jhu.edu}
}

\maketitle

\begin{abstract}
   Under tight feature and compute budgets, classical threat detection pipelines often degrade on near-decision-boundary events. Small quantum processors are now available, but existing work inadequately shows whether quantum components improve end-to-end threat detection under the above resource constrained conditions. This paper tries to address this gap with a hybrid architecture that uses a compact multilayer perceptron layer to compress the information in data and then routes the processed features to a few qubit quantum heads implemented in quantum support vector machine (QSVM) and variational quantum circuit (VQC) models. On a simulation platform, we benchmark these hybrid models against classical models with comparable parameter budgets on two representative cybersecurity tasks, network intrusion detection on NSL-KDD dataset and spam filtering on Ling-Spam dataset. To validate their precision on real quantum hardware, we deploy the best 4-qubit QSVM model on an IBM Quantum device with noise-aware execution, evaluated on a smaller sub-dataset. In the results, shallow quantum heads consistently match, and on difficult near-boundary cases modestly reduce missed attacks and false alarms compared to classical models using the same features. Hardware validation results track the simulation behavior closely enough that the remaining gap is dominated by device noise rather than model design. Furthermore, we conduct adversarial attacks to test the robustness of one QSVM model. Taken together, the study shows that even on small, noisy devices, carefully engineered quantum components may function as competitive, budget-aware components in practical cyber threat detection applications.
\end{abstract}


\section{Introduction}
\begin{comment}
\subsection{Security problem: detection under drift and tight budgets}
\label{sec:intro:security-problem}

Threat detection is deployed in an \emph{open} and \emph{non-stationary} environment: network workloads evolve, defenses change, and adversaries adapt, so dataset shift / concept drift is the norm rather than an exception~\cite{Shyaa2024ConceptDrift,Gama2014Survey,Quinonero2009DatasetShift}. Meanwhile, operational IDS must obey strict \emph{feature and compute budgets} (telemetry cost, latency SLAs, and limited labeling), which compresses representations and reduces redundancy, making decision boundaries more fragile under shift~\cite{Sommer2010ClosedWorld}. Finally, extreme class imbalance and asymmetric costs mean that small score changes can induce large swings in false alarms vs.\ missed attacks (the base-rate effect), so accuracy alone is insufficient; we must track FP/FN behavior under realistic operating points~\cite{Axelsson2000BaseRate,Davis2006PRROC}. This motivates our framing: under a fixed compact feature interface, can a decision-layer module reduce costly FP/FN---especially on borderline, hard-to-separate cases---while remaining stable under realistic constraints?
\end{comment}

\subsection{Motivation}
\label{sec:intro:why-quantum}

\begin{comment}
    
Cloud-accessible noisy intermediate-scale quantum (NISQ) processors are now usable. This study treats quantum machine learning (QML) as a \emph{budgeted decision-layer primitive} inside a classical security pipeline. It examines whether small ($2$--$4$ qubit) quantum heads can provide \emph{operationally meaningful} error reductions under strict resource constrains~\cite{Preskill2018QMLinNISQ,Cerezo2022Challenges}.

Quantum kernel methods as in quantum support vector machine (QSVM) realize a circuit-defined similarity geometry, and variational quantum classifiers (VQCs) implement a compact nonlinear decision layer via shallow parameterized circuits and interference/entanglement~\cite{Havlicek2019QuantumFeatureSpaces,Biamonte2017QML,Huang2021PowerOfData,PerezSalinas2020DataReupload}.

\end{comment}

Quantum machine learning (QML) for intrusion detection systems (IDS) has progressed from early proof-of-concepts to more system-style evaluations. Cloud-accessible noisy intermediate-scale quantum (NISQ) processors are now usable. Representative studies include QSVM-based NIDS prototypes on standard datasets often emphasizing feasibility and aggregate metrics ~\cite{Gouveia2020NetworkID,Kalinin2022SecurityID,Kim2024OutlierAnalysis}, and a recent work that benchmarks several QML heads, e.g.,  QSVM, VQC, and QCNN, and even runs on IBM backends with configurable resilience/decoupling~\cite{Abreu2024QMLIDS}.

While these results collectively suggest that QML can be competitive for cybersecurity attack detection, they also leave two gaps:

\begin{itemize}
    \item Attribution of performance improvement is often unclear. Many comparisons done mix different feature pipelines, dimensionality reductions, model capacities, and training budgets across classical machine learning baselines and quantum model variant choices~\cite{Gouveia2020NetworkID,Abreu2024QMLIDS}. 
    \item Noise-aware and security-centric evaluation is under-specified. The simulator--hardware gap is rarely characterized with a reproducible protocol that isolates shot noise, readout bias, and circuit-level noise while tracking false positive (FP) and false negative (FN) shifts under fixed operating points~\cite{Preskill2018QMLinNISQ,Cerezo2022Challenges,Nation2021M3}.
\end{itemize}

\begin{comment}
    
%% The following paragraph is repeated later multiple times.

In this work, we enforce a shared compact encoder and matched preprocessing/training budgets, a fixed q-qubit interface with identical preprocessing/splits and operating-point selection, so that performance differences are attributable to the quantum decision layer. Furthermore, we provide a concrete simulator$\rightarrow$IBM-hardware evaluation protocol and report the resulting FP/FN shifts and hard-case behavior under the same budgeted interface.
\end{comment}

\subsection{Research Questions and Tasks}
In this paper, we ask three questions:
\begin{itemize}
    \item Do quantum decision layers induce systematically different margin distributions, especially improving the low-margin regime where false alarms and misses are operationally costly?
    \item When NISQ execution noise is modeled as a kernel/estimator perturbation, can we determine whether the simulator–hardware gap remains within an acceptable range, as controlled by the perturbation magnitude and the decision margin?
    \item How stable is the quantum decision layer to evasion-style input perturbations? Can we characterize robustness in terms of worst-case performance drop over a family of bounded attacks, and relate the observed failure modes to margin/sensitivity of the induced kernel decision rule?
\end{itemize}
To answer these questions, we build a matched classical quantum pipeline that fixes preprocessing, the MLP encoder, and the q-feature interface, and varies only the decision layer: a matched linear classical head, a fidelity-kernel QSVM, or a shallow VQC with data re-uploading. We run a 2×2 factorial study over qubit count (q∈{2,4}) and quantum head type (QSVM vs. VQC) on NSL-KDD and Ling-Spam under shot-based noisy simulation. We then validate the best 4-qubit QSVM on IBM Quantum with M3 readout mitigation and XY8 dynamical decoupling. Furthermore, for spam detection, we ran a number of adversarial attacks perturbing email content and features to see how a selected QSVM model holds up in its detection performance. 

\subsection{Initial Findings}
This study shows that, under matched budgets, QSVM benefits materially from 2→4 qubits, while VQC is more optimizer/noise sensitive and can degrade at q=4 on NSL-KDD. On IBM hardware, the 4-qubit QSVM model achieves similar performance on a stratified 100-sample subset with mitigation/decoupling enabled, indicating a systematic but partially mitigable simulator–hardware gap under real device constraints. Finally, adversarial stress tests on the Ling-Spam QSVM pipeline show small drops under several perturbations but larger degradation under word-substitution and spam-noise attacks, with distinct FP/FN failure modes.

\section{Operational Setting and Threat Model}
\label{sec:threat_model}
\begin{comment}

\subsection{Defender's Goal}
 The defender aims to reduce both FP and FN, with emphasis on borderline/hard cases where operational cost concentrates.
 
\end{comment}

\subsection{Operational Assumptions}
\begin{itemize}
\begin{comment}
%% It is not clear the statement addresses the issue of concept drift.      
  \item Drift / non-stationarity: The data distribution may shift over time; we train with a fixed protocol and select operating thresholds on a held-out validation set.

\end{comment}
  
  \item Feature budget: The quantum head only consumes $q$ features ( $q\in\{2,4\}$). A compact classical encoder produces a normalized embedding, of which $q$ coordinates are routed to the quantum module.
  \item Compute budget: Circuit depth and shots are bounded. Kernel construction is performed with a bounded reference/training set (and/or batched/offline computation) to respect latency.
  \item NISQ noise: Hardware execution includes gate/measurement noise, finite-shot variance, and calibration drift. We allow evaluation-time mitigation (e.g., readout mitigation) and noise suppression (e.g., dynamical decoupling), but do not assume fault tolerance.
\end{itemize}

In the angle-encoding interface, one feature is mapped to one qubit rotation. Thus, q is both the feature-interface width and the number of qubits used by the quantum head.
\subsection{Adversarial Considerations}
\label{sec:adversary_model}
An initial investigation of test-time evasion via a stress test on Ling-Spam performs on a hybrid encoder + QSVM pipeline. It intends to gain insights of potential attack modes in the text pipeline and motivate deeper follow-up studies.

\begin{itemize}
    \item Attacker goal and knowledge: The attacker seeks to induce misclassifications, i.e., increased FN or FP, under black-/gray-box access. A white-box setting is used as a stress test via approximate gradients through the classical front-end.
    \item Attacker capability: The attacker perturbs emails under plausibility/readability constraints using token/phrase injection e.g., ``magic words''\cite{Cheng2022}, \cite{Tang2025}, character-level perturbations, word substitutions, spam-style noise injection, and combined attacks. We also consider three targeting modes: \texttt{ham\_only}, \texttt{spam\_only}, and \texttt{all}.
    \item Robustness characterization: We report post-attack drops in accuracy and confusion-matrix shifts to distinguish FP- vs.\ FN-dominant failures.
\end{itemize}

\section{Methodology}
\label{sec:method}
\begin{comment}
    
Our goal is to evaluate QML as a \emph{budgeted} decision-layer primitive for threat detection under tight feature and execution constraints. We therefore design an end-to-end pipeline where the \textbf{only} variable is the final decision module (classical vs.\ quantum), enabling clear attribution.

\end{comment}

\subsection{End-to-End Pipeline Overview}
The data-to-encoder pipeline with a classical or quantum head takes as input a preprocessed feature vector $x$. A compact MLP encoder $f_{\phi}$ outputs a $q$-dimensional embedding $\mathrm{u = f_{\phi}(x) \in \mathbb{R}^{q}}$ and $\mathrm{\qquad \|u\|_2 = 1}$, shared by:
\begin{itemize}
    \item Classical head: A lightweight linear/logistic classifier consumes the same $u$: $\mathrm{\hat{y} = \sigma(w^\top u + b)}$, serving as the matched classical baseline under the same feature budget.
    \item Quantum head: For quantum models, we map $u$ to rotation angles (with optional clipping for numerical robustness) $\mathrm{\theta_i = \pi \, u_i, where\quad i=1,\dots,q,}$ and feed $\theta$ into either a quantum kernel model (QSVM) or a trainable variational circuit (VQC).
\end{itemize}

To maintain matched budgets and enable attribution, preprocessing, training, validation, and test partitions, the MLP encoder, and the $q$-dimensional interface are held fixed across all methods. Resource assumptions are also matched (feature budget $q$, shot budget, and circuit depth constraints), so differences in detection performance can be attributed to the decision layer rather than upstream representation learning or data handling.

\begin{figure*}
    \centering
    \includegraphics[width=0.6\linewidth]{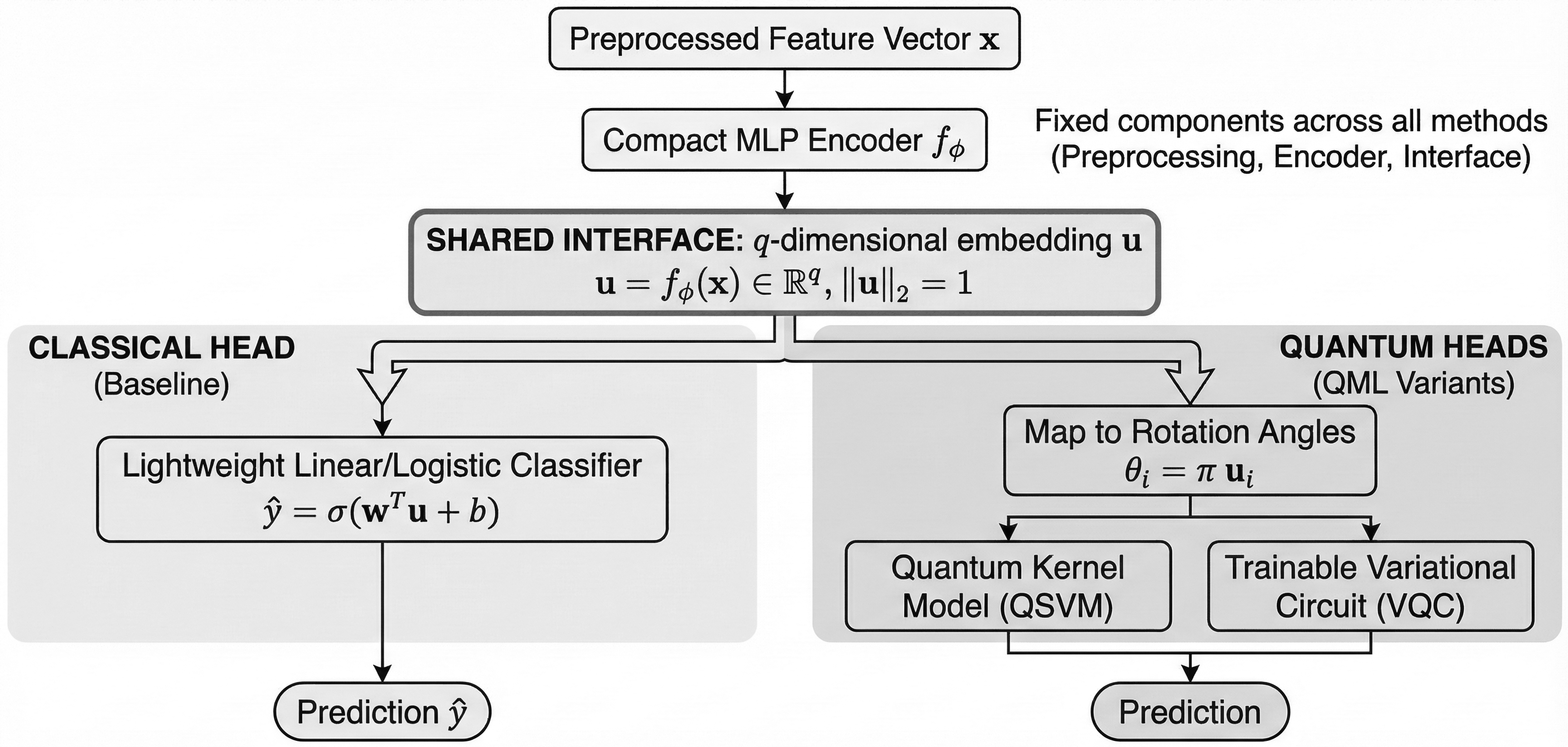}
    \caption{End-to-end pipeline overview}
    \label{fig:pipeline-overview}
\end{figure*}

\subsection{MLP Encoder}
The encoder directly outputs a $q$-dimensional, unit-norm embedding $u$, yielding a stable bounded interface for angle-based quantum encoding. We intentionally de-emphasize architectural complexity: the key design requirement is to produce a compact embedding that is (i) low-latency, (ii) stable on small/medium datasets, and (iii) shared across classical and quantum variants to preserve attribution.

\begin{comment}
\begin{figure*}
    \centering    
    \includegraphics[width=0.6\linewidth]{MLP.png}
    \caption{MLP Structure}
    \label{fig:mlp-structure}
\end{figure*}
\end{comment}

\subsection{Quantum Head A: QSVM}
\label{sec:method_qsvm}

The QSVM is a standard SVM whose kernel function is computed by a small quantum circuit. We only need a pairwise similarity score $k(u,v)$ between two $q$-dimensional embeddings $u,v \in \mathbb{R}^q$ from the MLP encoder.

Given $u$, we prepare a $q$-qubit quantum state $|\phi(u)\rangle$ using a fixed \texttt{ZZFeatureMap}. Intuitively, the circuit (i) applies feature-dependent single-qubit rotations so each feature influences one qubit, and (ii) uses ZZ-type entangling interactions so the state also depends on \emph{pairwise feature correlations}. The feature-map depth $r$ controls how much nonlinearity/correlation mixing is introduced while keeping circuits shallow.

We define the kernel between two inputs as the overlap between their feature-map states: $\mathrm{k(u,v) \;=\; \big|\langle \phi(u)\,|\,\phi(v)\rangle\big|^2 \in [0,1]}$. This quantity is a well-behaved similarity measure: it is 1 when the states match and near 0 when they are nearly orthogonal.

On real devices, we estimate $k(u,v)$ by a simple procedure that avoids full state tomography:
\begin{enumerate}
  \item Prepare $|\phi(u)\rangle$ by running the feature map on $|0\rangle^{\otimes q}$.
  \item Apply the inverse feature map for $v$, i.e., $U_\phi(v)^\dagger$.
  \item Measure in the computational basis; the probability of observing the all-zero bitstring approximates $|\langle \phi(u)|\phi(v)\rangle|^2$.
\end{enumerate}
With $S$ shots, each kernel entry is estimated from empirical frequencies. This makes the QSVM kernel explicitly \emph{noise- and shot-aware}: estimation variance and hardware errors translate into controlled perturbations of the Gram matrix.

For a training set of size $n$, we build the Gram matrix $K \in \mathbb{R}^{n\times n}$ with entries $K_{ij}=k(u_i,u_j)$. This requires $O(n^2)$ kernel evaluations (and thus $O(n^2 S)$ shots on hardware), which is the dominant cost.

Kernel methods are sensitive to global mean shifts in feature space (especially under drift or sampling imbalance). To ensure consistent geometry and avoid biasing the margin by the kernel mean, we apply standard double-centering on the \emph{training} kernel.

\begin{comment}
    
\begin{equation}
H \;=\; I - \frac{1}{n}\mathbf{1}\mathbf{1}^\top,\qquad
K_c \;=\; H K H.
\end{equation}

\end{comment}

For validation and test samples, we apply the corresponding out-of-sample centering using the training statistics. This step makes the QSVM decision function depend on relative similarities rather than absolute offsets that can vary across partitions.

We train a classical SVM using the centered precomputed kernel. At inference time, prediction for a sample $u$ requires computing its kernel vector against the training (or reference) set. The only quantum component is computing $k(u,\cdot)$.

\begin{figure*}
    \centering    
    \includegraphics[width=0.6\linewidth]{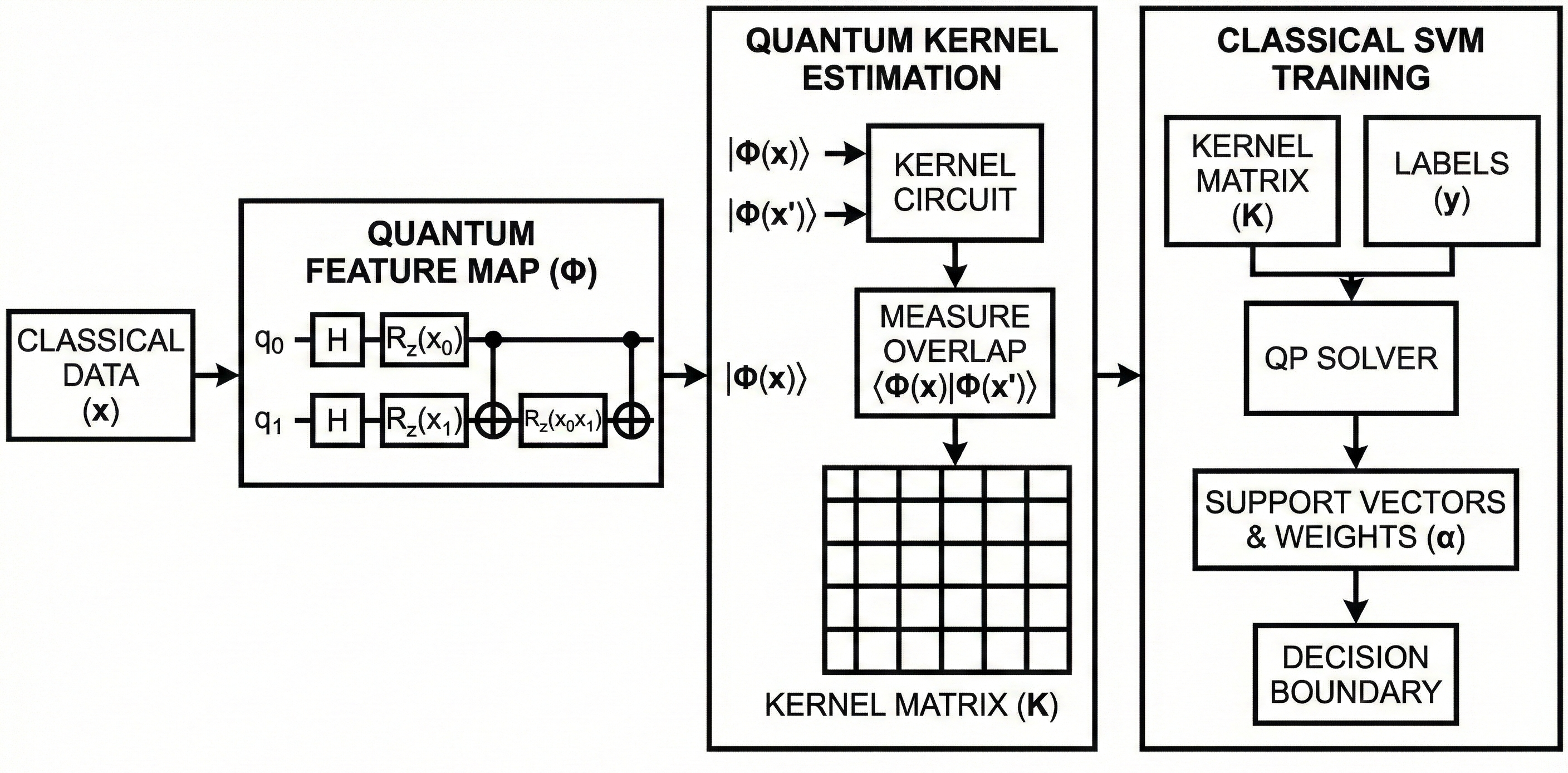}
    \caption{QSVM Pipeline}
    \label{fig:qsvm-pipeline}
\end{figure*}

\subsection{Quantum Head B: VQC}
\label{sec:method_vqc}

While QSVM uses a \emph{fixed} feature map and learns only the SVM weights, the VQC learns a \emph{trainable} quantum circuit end-to-end: circuit parameters are optimized to separate benign vs.\ malicious samples under the same $q$-feature interface. Given $u \in \mathbb{R}^q$, we construct a circuit with $L$ re-uploading blocks: $\mathrm{U(u;\omega) \;=\; \prod_{\ell=1}^{L} \Big( U_{\text{ans}}(\omega_\ell)\,U_{\text{feat}}(u) \Big)}$, where $U_{\text{feat}}(u)$ encodes features as rotation angles (the same $\theta_i=\pi u_i$ interface used by QSVM), and $U_{\text{ans}}(\omega_\ell)$ is a shallow \texttt{TwoLocal}-style ansatz layer.
Concretely, each ansatz layer applies parameterized single-qubit rotations on all qubits followed by a sparse entangling pattern (e.g., linear/ring nearest-neighbor entanglement). This design keeps depth low and respects NISQ constraints while still enabling feature interaction through entanglement.

With very few qubits, representational capacity is limited if we encode $u$ only once. \emph{Data re-uploading} addresses this by injecting the same $q$ features multiple times, interleaved with trainable mixing layers. Intuitively, this lets the circuit build richer nonlinear decision boundaries \emph{without increasing qubit count}, trading a small amount of depth for expressivity in a controlled, hardware-compatible way.

After the circuit, we measure a simple observable readout (e.g., $Z$ on one qubit or a fixed Pauli string) and use its expectation as a logit: $\mathrm{z(u) \;=\; \langle Z \rangle_{U(u;\omega)}}$ and $\mathrm{\qquad \hat{y} \;=\; \sigma(\alpha z(u)+\beta)}$. With finite shots, $z(u)$ is estimated from measurement frequencies, making the classifier explicitly stochastic at train time.

We train the VQC parameters $\omega$ (and optionally the affine calibration $\alpha,\beta$) using a standard binary classification loss (cross-entropy or BCE-with-logits) on the same training and validation partitions as all other variants. The two practical design choices were:
\begin{itemize}
  \item Shot noise and optimization noise are part of the model. Unlike deterministic classical logits, $z(u)$ is estimated from samples, so we keep circuits shallow and use validation-loss monitoring for early stopping.
  \item Expressivity vs.\ trainability is a first-class trade-off. Deeper/more-parameter circuits can fit better but are more sensitive to barren plateaus and hardware noise. We therefore restrict depth (via \texttt{TwoLocal} and small $L$) and treat $L$ as a controlled knob rather than maximizing parameters.
\end{itemize}

Methodologically, the VQC tests a different hypothesis than QSVM: whether a \emph{learnable} quantum decision layer can adapt to hard/near-boundary cases under the same strict $q$-feature interface. Empirically, it is ``higher upside but higher risk'' than QSVM.

\begin{figure*}
    \centering    
    \includegraphics[width=0.80\linewidth]{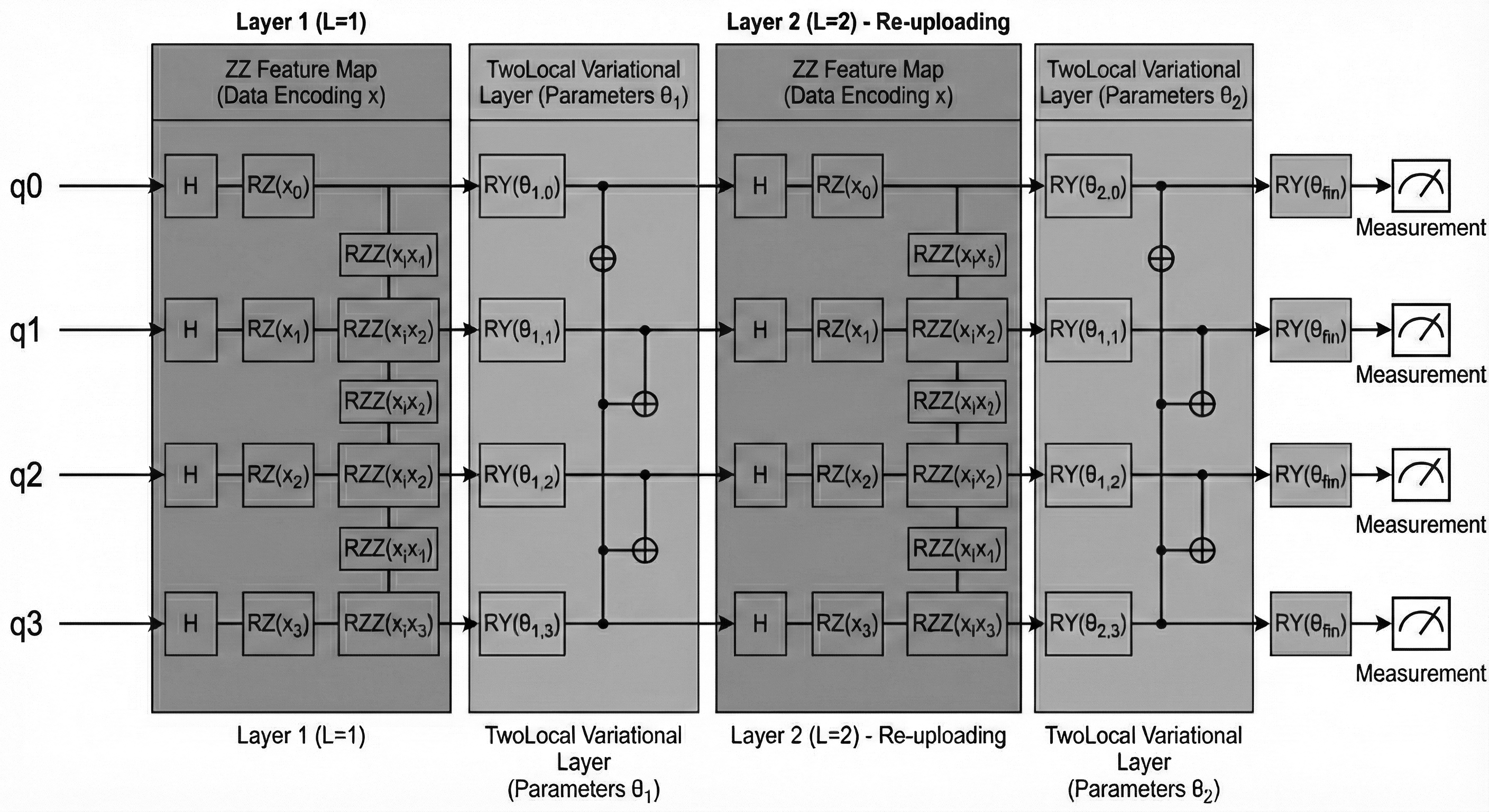}
    \caption{VQC Structure}
    \label{fig:vqc-structure}
\end{figure*}

\subsection{Hardware Execution}
All model fitting and selection tasks are completed before hardware execution. We fit the MLP encoder and QSVM on the training partition using simulated kernels, then use the validation partition to select the 4-qubit configuration and operating threshold. The preprocessor, encoder, feature map, SVM, reference set, kernel-centering statistics, and threshold are frozen before final evaluation.

For validation on hardware, kernel fidelities for a stratified 100-sample subset of the held-out test set are estimated on an IBM Quantum backend. We use Qiskit Runtime Sampler and batched compute--uncompute circuits to estimate the probability required for $k(u,v)$. M3 readout mitigation and XY8 dynamical decoupling are enabled only during this hardware evaluation. Hardware measurements form test kernel vectors but do not update learned parameters or influence model selection.

\section{Experimental Design}
\label{sec:exp_design}

\begin{comment}

Our evaluation is structured as a \emph{controlled security-systems protocol} with an attribution goal: we hold data handling, splits, and feature/compute budgets fixed, and vary only the decision layer so observed differences can be credibly attributed to the (classical vs.\ quantum) head.

\end{comment}

\subsection{Datasets and Preprocessing}
\label{sec:datasets_preproc}

For intrusion detection, we divide \texttt{KDDTrain+.txt} into stratified training and validation partitions and use \texttt{KDDTest+.txt} as the held-out test set. To prevent leakage, all preprocessors are fit on the training partition and then reused without refitting for validation and test samples.
Numeric features are standardized; categorical fields are rare-bucketed (minimum count threshold) and one-hot encoded. Class imbalance is handled using class-weighted training (balanced weights).

For spam detection, we load emails from the \texttt{lingspam\_public} tree (e.g., \texttt{spmsg*} $\rightarrow$ spam) and create stratified training, validation, and test partitions.
Text is vectorized using TF--IDF with standard tokenization choices (lowercasing, English stopwords, document-frequency filters; $n$-grams per configuration).
We again apply balanced class weights to address imbalance.

For both datasets, the training partition is used to fit all the preprocessors, the MLP encoder, and the decision heads. The validation partition is used only for early stopping, configuration and threshold selection. The test partition remains held out until the final simulator, hardware, and robustness evaluations.

\begin{comment}

NSL-KDD has considerable age and realism limitations. In our experiments, we use it as a controlled, auditable baseline to validate protocol correctness (leakage control, matched budgets, and simulator-to-hardware transfer), and our future work expands to more modern, drift-relevant datasets.

\end{comment}

\subsection{Classical Baselines}
\label{sec:baselines}

We include \emph{parameter-light, budget-matched} classical baselines designed to be comparable to quantum heads under the same feature interface and with the same threshold/operating-point selection protocol.

\begin{comment}

Across all baselines and quantum variants, we enforce:
\begin{itemize}
  \item identical preprocessing and splits;
  \item the same feature budget (only the first $q$ embedding coordinates are routed to the head);
  \item the same threshold/operating-point selection protocol.
\end{itemize}
This ensures differences remain attributable to the decision layer rather than upstream handling.
\end{comment}

\subsection{Factorial Controls and Reproducibility Knobs}
\label{sec:repro_controls}

All quantum-circuit simulations are implemented in Qiskit and executed with the shot-based Qiskit Aer simulator. Each circuit uses 1,024 shots and a depolarizing noise model with single- and two-qubit error probabilities of $10^{-3}$ and $10^{-2}$, respectively.

We run a $2\times2$ factorial design over qubit count ($q\in\{2,4\}$) and QML head (QSVM or VQC). All non-factor components are held fixed, including the shared MLP encoder, feature interface, data splits, and evaluation metrics. The shared encoder is trained for at most 25 epochs with a batch size of 256, a learning rate of $10^{-3}$, weight decay of $10^{-4}$, dropout of 0.15, early-stopping patience of 8, and balanced class weights.

After selecting the configuration and operating threshold on the validation set, we evaluate the frozen 4-qubit MLP+QSVM on IBM Quantum using M3 readout mitigation and XY8 dynamical decoupling. Because of the hardware budget, this evaluation uses a stratified 100-sample subset of the held-out test set and records per-sample outputs for auditability.

The hardware evaluation protocol employs the following steps:
\begin{enumerate}
    \item compute the $q$-dimensional MLP embedding and map it to rotation angles,
    \item build one kernel row against a bounded reference set by running batched compute--uncompute circuits and extracting $p(0^q)$ per pair,
    \item center the kernel row using statistics computed from the training kernel,
    \item run the pre-trained SVM decision function, and
    \item stream per-sample outputs while updating the running confusion matrix (TP/TN/FP/FN) for auditability.
\end{enumerate}

\subsection{Cost and Scalability}
The main deployment barrier is cost/latency predictability: QSVM kernel construction scales quadratically with the reference set size, and finite shots plus hardware queue/drift introduce variance.

For practicality, we employ the following strategy to:
\begin{enumerate}
    \item cap and stratify the reference set ($n_{\text{ref}}$) and cache kernel rows,
    \item use Nystr\"om / randomized low-rank approximations to reduce $O(n^2)$ kernel work to $O(n n_{\text{ref}})$ while preserving the attribution principle,
    \item adopt margin-aware shot allocation (more shots only when decision margins are small), and
    \item run the quantum kernel asynchronously/offline with periodic refresh to mitigate queue time and calibration drift.
\end{enumerate}

\section{Results}

\subsection{Evaluation Metrics}
\label{sec:metrics}

Macro-F1 is the unweighted mean of the class-specific F1 scores, $\mathrm{(F1_0+F1_1)/2}$, where each F1 score is the harmonic mean of precision and recall. The area under the receiver operating characteristic curve (AUROC) measures ranking performance across classification thresholds. The area under the precision--recall curve (AUPRC) emphasizes performance on the positive class and is particularly informative for imbalanced security datasets.

\subsection{Matched-Budget Performance}
For intrusion detection, as in Table \ref{tab:nslkdd_results}, the 4-qubit QSVM achieves the best overall balance (highest macro-F1 and competitive AUROC/AUPRC). Increasing qubits from $q{=}2$ to $q{=}4$ yields a large gain for QSVM, consistent with the hypothesis that additional qubits expand the effective feature-map capacity. In contrast, the VQC is less stable: while it can reach high AUROC/AUPRC, it degrades in macro-F1 at $q{=}4$, suggesting optimizer/noise sensitivity that is operationally relevant when false alarms vs.\ misses must remain balanced.

\begin{table}[!htbp]
\centering
\caption{Results on NSL-KDD under shot-based noisy simulation}
\label{tab:nslkdd_results}
\begin{tabular}{lcccc}
\toprule
Model & Acc & F1$_{\text{macro}}$ & AUROC & AUPRC \\
\midrule
MLP+Linear & 0.6911 & 0.6613 & 0.7411 & 0.7032 \\
MLP+QSVM ($q{=}2$)            & 0.8410 & 0.8322 & 0.8802 & 0.9057 \\
MLP+QSVM ($q{=}4$)            & \textbf{0.9212} & \textbf{0.9098} & 0.9335 & 0.9407 \\
MLP+VQC ($q{=}2$)             & 0.8522 & 0.8410 & 0.8899 & 0.9289 \\
MLP+VQC ($q{=}4$)             & 0.8073 & 0.7907 & \textbf{0.9365} & 0.9402 \\
\bottomrule
\end{tabular}
\end{table}

For spam detection, as in Table \ref{tab:lingspam_results}, all hybrid heads are strong. The QSVM models remain the most consistent across metrics under the same feature budget. The four-qubit QSVM reaches near-ceiling performance across all metrics.

\begin{table}[!htbp]
\centering
\caption{Results on Ling-Spam under shot-based noisy simulation}
\label{tab:lingspam_results}
\begin{tabular}{lcccc}
\toprule
Model & Acc & F1$_{\text{macro}}$ & AUROC & AUPRC \\
\midrule
MLP+Linear & 0.9214 & 0.9001 & 0.9504 & 0.9103 \\
MLP+QSVM ($q{=}2$)            & 0.9883 & 0.9710 & 0.9990 & 0.9768 \\
MLP+QSVM ($q{=}4$)            & \textbf{0.9996} & \textbf{0.9987} & \textbf{1.0000} & \textbf{1.0000} \\
MLP+VQC ($q{=}2$)             & 0.9810 & 0.9788 & 0.9997 & 0.9894 \\
MLP+VQC ($q{=}4$)             & 0.9857 & 0.9552 & 0.9995 & 0.9985 \\
\bottomrule
\end{tabular}
\end{table}

\subsection{Hardware Evaluation}

Table~\ref{tab:hw_results} reports the final hardware-evaluation results for the frozen 4-qubit MLP+QSVM on the stratified 100-sample test subset. With M3 readout mitigation and XY8 dynamical decoupling enabled, Ling-Spam remained near ceiling, whereas NSL-KDD showed a larger decrease relative to simulation. These differences are consistent with device noise and execution constraints, but the small hardware sample does not isolate their individual contributions.

\begin{table}[H]
\centering
\caption{IBM Quantum evaluation of the 4-qubit MLP+QSVM}
\label{tab:hw_results}
\begin{tabular}{lcccc}
\toprule
Dataset & Acc & F1$_{\text{macro}}$ & AUROC & AUPRC \\
\midrule
NSL-KDD   & 0.8400 & 0.8242 & 0.8924 & 0.9602 \\
Ling-Spam & 0.9900 & 0.9827 & 0.9986 & 0.9738 \\
\bottomrule
\end{tabular}
\end{table}

\subsection{Robustness Stress Tests}

We stress-test the frozen 4-qubit MLP+QSVM on the held-out Ling-Spam test set ($n=2{,}315$). Metrics are computed on the same samples, while \texttt{spam\_only}, \texttt{ham\_only}, and \texttt{all} perturb spam, ham, or both classes, respectively. Each drop is defined as $\Delta=\text{clean}-\text{adversarial}$.

Under \texttt{spam\_only}, magic-word injection adds 50 trigger tokens, while character perturbation applies a swap, insertion, deletion, or duplication with probability 0.15 per token. FGSM ($\epsilon=0.5$, one step) and PGD ($\epsilon=0.8$, 10 steps) perturb the continuous classical feature representation before the quantum head. Under \texttt{ham\_only}, word substitution replaces tokens with probability 0.2 using a target vocabulary, while high-intensity spam noise inserts short spam-like phrases at randomized positions. Under \texttt{all}, the combined attack applies these four text-level operators to both classes using the same parameters.

Table~\ref{tab:robustness} uses clean baselines of 0.9996 accuracy and 0.9975 spam-class F1. The combined attack yields the largest drops (0.0372 and 0.0656). Spam noise and word substitution produce the next-largest F1 drops, 0.0540 and 0.0450, respectively; all remaining F1 drops are at most 0.0381. These results are preliminary stress tests rather than certified robustness guarantees.

\begin{table}[!htbp]
%%\begin{table*}
\centering
\caption{Robustness stress tests of the 4-qubit Ling-Spam MLP+QSVM, grouped by targeting mode}
\label{tab:robustness}
%%\scriptsize
\begin{tabular}{lrrrr}
\toprule
Attack (params) & Acc$_{\text{adv}}$ & $\Delta$Acc & F1$_{\text{spam,adv}}$ & $\Delta$F1 \\
\midrule
\multicolumn{5}{l}{\texttt{spam\_only} (spam samples perturbed)} \\
magic\_words(50)                       & 0.9888 & 0.0108 & 0.9640 & 0.0335 \\
char\_perturbation(0.15)               & 0.9883 & 0.0113 & 0.9638 & 0.0337 \\
FGSM($\epsilon{=}0.5$, 1 step) & 0.9875 & 0.0121 & 0.9623 & 0.0352 \\
PGD($\epsilon{=}0.8$, 10 steps) & 0.9801 & 0.0195 & 0.9594 & 0.0381 \\
\midrule
\multicolumn{5}{l}{\texttt{ham\_only} (ham samples perturbed)} \\
word\_substitution(0.2)                & 0.9823 & 0.0173 & 0.9525 & 0.0450 \\
spam\_noise(high)                      & 0.9814 & 0.0182 & 0.9435 & 0.0540 \\
\midrule
\multicolumn{5}{l}{\texttt{all} (both classes perturbed)} \\
combined (all samples)                 & 0.9624 & 0.0372 & 0.9319 & 0.0656 \\
\bottomrule
\end{tabular}
\end{table}

\section{Discussion}
The main observations from this study are: (i) a 4-qubit QSVM is the most consistent performer across tasks; (ii) on-device validation preserves the simulator ranking with a systematic but explainable simulator$\rightarrow$hardware gap; (iii) preliminary robustness stress tests show that failures concentrate in specific, security-relevant perturbation families rather than in superficial edits.

Our results suggest that small, NISQ-era quantum decision layers can be \emph{operationally credible} components in budgeted threat-detection pipelines, provided the evaluation is attribution-safe (shared preprocessing/encoder/interface) and hardware effects are treated as first-class constraints rather than an afterthought.

Under matched budgets, the quantum-kernel head (QSVM) is consistently the most stable variant and benefits materially from increasing the interface width from $q{=}2$ to $q{=}4$ qubits on both datasets. In contrast, the variational head (VQC) is less predictable: additional qubits increase expressivity but also amplify optimizer sensitivity and exposure to two-qubit noise, which can negate gains (especially on tabular IDS).

Hardware validation of the best $4$-qubit QSVM shows that the simulator trend transfers, but with a systematic degradation consistent with three mechanisms: (i) device ISA/coupling constraints increasing effective depth (routing/SWAPs), (ii) finite-shot estimator variance that matters most near the decision boundary, and (iii) systematic bias from readout/crosstalk and calibration drift.

The Ling-Spam stress tests indicate that some plausible perturbations cause only modest degradation, but word-substitution and spam-noise attacks induce larger drops with distinct FP vs.\ FN failure modes. This points to the encoder$\rightarrow$quantum interface (tokenization/TF--IDF, truncation, normalization) as security-critical: improving robustness requires mechanisms in the classical frontend and evaluation protocols (attack budgets, semantic constraints), in addition to the quantum head.

\begin{comment}

This study is intentionally budgeted: hardware validation uses a small stratified subset; hyperparameter search is narrow; and drift is only indirectly modeled via held-out splits. The most actionable follow-ups are:
\begin{enumerate}
    \item repeat hardware runs across multiple calibration windows and report variance,
    \item rerun simulation with device-constrained bases/coupling and calibration-informed noise models to attribute residual transfer gaps,
    \item evaluate under explicit drift (temporal splits, cross-regime testing) using operating-point metrics (e.g., TPR at fixed FPR), and 
    \item reduce QSVM kernel cost via bounded reference sets / Nystr\"om-style approximations while preserving leakage-safe attribution.
\end{enumerate}

\end{comment}

This work establishes an attribution-safe evaluation pipeline and an initial on-device validation. The next steps in follow-up studies should strengthen statistical evidence, evaluate operational realities (drift and adversaries), and clarify when a quantum decision layer is worth deploying under tight budgets.

\bibliographystyle{IEEEtran}
\bibliography{references}


\end{document}